\documentclass[aps,prb,twocolumn,superscriptaddress,showpacs,floatfix]{revtex4}
\usepackage{graphicx}
\usepackage{latexsym}
\usepackage{stmaryrd}
\usepackage{amsmath}
\begin{document}
\title{Quantum anomalous Hall effect with tunable Chern number in magnetic topological insulator film}
\author{Hua Jiang}\email{jianghuaphy@gmail.com}
\affiliation{International Center for Quantum Materials, Peking University, Beijing 100871, China}
\author{Zhenhua Qiao}
\affiliation{Department of Physics, The University of Texas at
Austin, Austin, Texas 78712, USA}
\author{Haiwen Liu}
\affiliation{Institute of Physics, Chinese Academy of Sciences, Beijing 100190, China}
\author{Qian Niu}
\affiliation{Department of Physics, The University of Texas at
Austin, Austin, Texas 78712, USA}\affiliation{International Center for Quantum Materials, Peking University, Beijing 100871, China}
\date{\today}

\begin{abstract}
We study the possibility of realizing quantum anomalous Hall effect (QAHE) with tunable Chern number through doping magnetic elements in the multi-layer topological insulator film. We find that high Chern number QAHE phases exist in the ideal neutral samples and it can make transition to another QAHE phase directly by means of tuning the exchange field strength or  sample thickness. With the help of an extended Haldane model, we demonstrate the physical mechanism of the tunable Chern number QAHE phase. Further, we study the phenomena in the realistic samples with the Fermi energy being gating-modulated, where the multiple-integer QAHE plateaus remain and two separated QAHE phase are connected by metallic phase. The stability of the high Chern number QAHE phase is also discussed.
\end{abstract}
\pacs{73.43.Nq, 72.15.Rn, 72.25.-b, 85.75.-d}
\maketitle

\section{Introduction}
The integer quantum Hall effect~\cite{Klitzing,DCTUSI} is one of the most important discoveries in  condensed matter physics. When a strong perpendicular magnetic field is applied to the two-dimensional electron gas at low temperatures, the Hall conductance exhibits a precise quantization in units of $e^2/h$ (i.e. the fundamental conductance unit) due to  Landau-level quantization. This quantization is directly connected to the topological properties of the two-dimensional bulk states, characterized by a topological invariant $\mathcal{C}$ known as the first Chern number.~\cite{RBLauglin,Thouless}

Non-zero Chern number can in principle also occur in the band structure of other systems with time-reversal symmetry breaking, such as in a ferromagnet, leading to the so-called quantum anomalous Hall effect(QAHE). This effect was first proposed in a seminal paper by Haldane~\cite{Haldane} in a honeycomb lattice model with an average flux per unit cell being zero. Due to its unique nontrivial topological properties and  great potential application for designing dissipationless spintronics, extensive studies have been made to search possible host materials of realizing such QAHE. Several candidate systems are proposed recently including the ferromagnetic Mercury based quantum wells in the insulating state,~\cite{CXLiu} the disorder induced Anderson insulator,~\cite{Nagaosa} Rashba graphene coupled with exchange field,~\cite{ZHQiao} Kagome lattice,~\cite{Nagaosa1,ZYZhang} and optical lattice models.~\cite{CJWu}
\begin{figure}
\includegraphics [width=\columnwidth, viewport=123 163 707 482, clip]{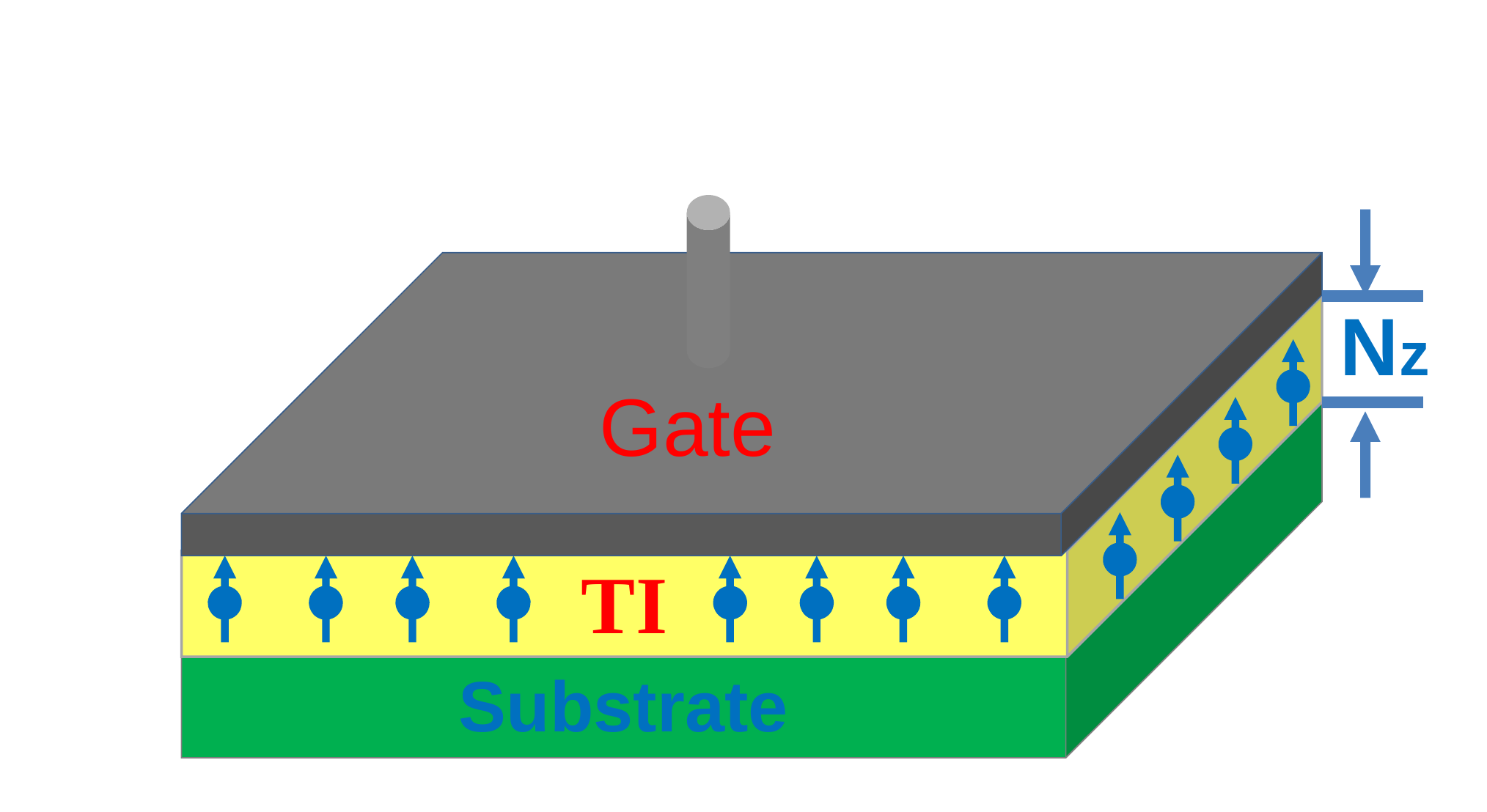}
\caption{(Color online)  Schematic plot of the setup: a magnetically doped topological insulator on a substrate with its Fermi energy being tuned by the attached external gate. Blue arrows indicate the magnetized dopants. $\rm N_z$ measures the sample layers along Z-direction.}\label{schematicPlot}
\end{figure}

However, although the QAHE has been theoretically proposed, the Chern numbers in the proposed QAHE systems can only take some low and limited values~(i.e. $\mathcal{C}=1,~2,~4$). This is distinct from that in the conventional integer quantum Hall effect, where the Chern number can be very high and tuned continuously through adjusting the external magnetic field or the Fermi energy. Therefore, a natural question arises: whether it is possible to find some host materials exhibiting QAHE with tunable Chern numbers? In the following, we shall give a positive answer by doping magnetic elements into the topological insulator films. Topological insulator is new quantum phase of matter that behaves as an insulator in the bulk but carries time-reversal symmetry protected odd-number pairs of helical edge states (two-dimensional) or surface states (three-dimensional).\cite{CLKane1}

The topological insulator was first proposed in two-dimensional materials, however, there are only limited candidates showing the theoretical possibility of hosting  two-dimensional topological insulator.~\cite{CLKane2,SCZhang1,SCZhang2,Fertig,ZHQiao1,RQWu,RRDu,YGYao} In contrast, there have been numerous materials or compounds found to host the three-dimensional topological insulators both theoretically ~\cite{LiangFu,HJZhang,Dixiao,ZhongFang,Yansun,Virot} and experimentally \cite{DHsieh,Xia,Sato}. This has attracted extensive attentions from the condensed-matter-physics community, and provides an interesting platform for the practical application in low-power dissipation spintronics. For instance, it has reported that magnetically doped topological insulator could exhibit the QAHE using  first principal theory,~\cite{RuiYu} and recent experimental discovery has shown the giant anomalous Hall conductance in the magnetic doped topological insulator, which is a good sign for the final realization of QAHE.~\cite{QKXue}

In this article, we show that the QAHE with tunable Chern numbers can be achieved through doping magnetic elements in a topological insulator film. The magnetic dopant in the topological insulator results in two effects:~\cite{ChangKai,YLChen} (a) breaking the time-reversal symmetry; (b) inducing an effective exchange field. Through investigating the evolution of the bulk band structure, counting the edge states winding number and the direct Hall conductance calculation, we determine the existence of QAHE phases with tunable Chern numbers in our studied system. Using an extended Haldane model, we give an explanation on the mechanism of the tunable Chern number QAHE phases. Furthermore, the influence of Fermi energy fixed by external gate are discussed. When the Fermi energy is slightly deviated from the zero energy, the high Chern number QAHE phases can still survive even though the metallic phases emerge between two separated QAHE phases. In the end, we study the disorder effect on the obtained QAHE phase and find that the high Chern number QAHE plateaus can be very stable against weak disorders.

The remaining of this paper is organized as follows. In Sec.~\ref{model}, we introduce an theoretical model and the methods for calculating kinds of topological features. In Sec.~\ref{HalfFilling}, we present the numerical results for the ideal neutral samples and discuss their mechanisms. In Sec.~\ref{RealisticSamples}, we show the influence of the Fermi energy shifting and disorder on tunable Chern number QAHE which are inevitably present in realistic samples. Finally, a brief discussion and conclusion are given in Sec.~\ref{Summary}.

\section{Model and methods}\label{model}
Figure~\ref{schematicPlot} plots the schematic setup: the topological insulator material is doped with magnetic elements, with the arrows representing the magnetic dopants. We use $\rm {N_Z}$ to label the thickness of the topological insulator layers. As a starting point, we introduce the effective Hamiltonian. In the cubic lattice model, the electronic state at each site ${\bf i}$ can be expressed as $\phi_{\bf i}=[a_{\bf i \uparrow},~b_{\bf i \uparrow},~a_{\bf i \downarrow},~b_{\bf i \downarrow}]^{T}$, where ($a,b$) denote two independent orbits and ($\uparrow,\downarrow$) represents spin indices. In the tight-binding representation, the topological insulator doped with magnetic elements can be written as:~\cite{HJZhang,SQShen,HMGuo}
\begin{eqnarray}
&& H= H_{3D} +H_{imp}   \nonumber\\
&& H_{3D}= \sum_{\bf i} E_0 \phi_{\bf i}^{+} \phi_{\bf i} + \sum_{\bf i}\sum_{\alpha= x,  y,  z} \phi_{\bf i}^{+} T_{ \alpha} \phi_{{\bf i}+\widehat{\alpha}} +h.c.   \nonumber\\
&& H_{imp} = \sum_{\bf i} m_{0} \phi_{\bf i}^{+} \phi_{\bf i}\label{equation1}
\end{eqnarray}
where $H_{3D}$ describes the bulk Hamiltonian of the three-dimensional topological insulator, $E_0$, $T_{ \alpha}$, $m_0$ are written as:
\begin{eqnarray}
&& E_{0}=(M-\sum_\alpha B_\alpha ) \sigma_0 \otimes \tau_z - \sum_\alpha D_\alpha \sigma_0 \otimes \tau_0 \nonumber\\
&& T_{ \alpha} = \frac{B_\alpha}{2} \sigma_0 \otimes \tau_z + \frac{D_{\alpha}}{2} \sigma_0 \otimes \tau_0 -\frac{i A_{\alpha}}{2} \sigma_\alpha \otimes \tau_x \nonumber\\
&& m_0 =m \sigma_z \otimes \tau_0
\end{eqnarray}
where $M$, $A_\alpha$, $B_\alpha$, $D_{\alpha}$ are independent parameters with M determining the inverted band gap amplitude and $A_{\alpha}$ reflecting the Fermi velocity. $\widehat{\alpha}$ is the unit vector along $\alpha=(x,y,z)$ direction. $\sigma$ and $\tau$ are the Pauli matrices in  spin and orbital spaces, respectively.  $H_{imp}$ is used to describe the magnetic dopants. $m$ measures the effective exchange field strength. For simplicity, we assume that the Lande $g$ factor is the same for all bases, which is reliable because in realistic three-dimensional topological insulator all lowest-energy bands determining the band topology are always combined by the p-orbitals~\cite{HJZhang} or d-orbitals. Since we focus on the resulting phenomena of the topological insulator film, the layers along the Z-direction are set to be finite.

In general, analyzing the evolution of the bulk and edge energy spectrum as functions of some tunable parameters is an efficient way to investigate the topological features of a system. Let us first focus on the bulk energy spectrum. Since the studied system has the translational
symmetry along both $x$ and $y$ directions, both the corresponding momenta $k_x$ and $k_y$ are good quantum numbers. Through performing the partial Fourier transformation
\begin{eqnarray}
\phi_{ k_x k_y} (z)=\frac{1}{\sqrt{L_x L_y}} \sum_{ x, y } e^{i k_x x + i k_y y } \phi_{\bf i} (x,y,z),
\end{eqnarray}
the real-space Hamiltonian in Eq.~(\ref{equation1}) becomes
\begin{eqnarray}
H_1(\bf{k}) &=&     \sum_{k_x k_y,z} \{ \phi_{k_x k_y}^{+} (z) [E_0+m_0] \phi_{k_x k_y} (z) \} \nonumber\\
&+& \sum_{k_x k_y,z} \{   \phi_{k_x k_y}^{+}(z) [T_{x} e^{i k_x}+ T_{y} e^{i k_y} ] \phi_{k_x k_y}(z) \nonumber\\
&+&  \phi_{k_x k_y}^{+} (z) T_z \phi_{k_x k_y}(z+1)  +h.c.\}.\label{Ham-K}
\end{eqnarray}
By directly diagonalizing the Hamiltonian in Eq.~(\ref{Ham-K}), one can obtain the bulk band spectrum.

In order to study the edge state physics, one has to consider a boundary or surface, i.e. in our case, we choose to terminate the topological insulator along $y$ direction and consequently an edge exists in the $y=0$ plane. Therefore, only $k_x$ is left to be a good quantum number. The corresponding partial Fourier transformation and the resulting momentum-space Hamiltonian $H_2(k_x)$ can be expressed as:
\begin{eqnarray}
&&\phi_{k_x}(y,z) =  \frac{1}{\sqrt{L_x}} \sum_{ x, y } e^{i k_x x } \phi_{\bf i} (x,y,z),  \nonumber\\
&&H_2(k_x) = \sum_{k_x,y,z}  \{ \phi_{k_x}^{+} (y,z) [E_0+m_0] \phi_{k_x} (y,z)\} +\sum_{k_x,y,z}  \nonumber\\
&& \{ \phi_{k_x}^{+} (y,z)   T_x e^{i k_x}  \phi_{k_x}(y,z) + \phi_{k_x}^{+} (y,z) T_z  \phi_{k_x}(y,z+1) \nonumber\\
&& +h.c. \} +\sum_{k_x,y,z} \{ \phi_{k_x}^{+} (y,z) T_z  \phi_{k_x}(y+1,z) +h.c. \}.  \nonumber\\
&&=\sum_{k_x} H_{1D} (k_x)
\end{eqnarray}
In this way, the system can be treated as a semi-infinite quasi-one dimensional tight-binding chain along $y$ direction. Using the nonequilibrium Green's  function technique, the edge Green's function $G^r(k_x,E,y=0)$ can be numerically obtained,~\cite{Lopez} where $E$ denotes the Fermi-energy. Both the bulk and edge energy spectra information can be included in the spectral function ${\bf {A}}(k_x,~E)$ of the edge, which reads
\begin{eqnarray}
{\bf {A}}(k_x,E)=-\frac{1}{\pi} {\rm {Im \{Tr}} [G^r(k_x,E,y=0)]\}.
\end{eqnarray}

In addition, the topological invariant is another important quantity to characterize the topological properties of the system. For example, in the absence of external magnetic field a better way to judge  QAHE is to see whether the Chern number is non-zero or not. The Chern number equals to the zero temperature Hall conductance and can be calculated via the Kubo-formula:~\cite{Thouless}
\begin{eqnarray}
\sigma_{xy}=\frac{e^2}{\hbar} \int \frac{d k_x d k_y}{(2 \pi)^2} \sum_{\varepsilon_{l} < E_F < \varepsilon_{n}} {\rm Im} \frac{\langle l| \frac{\partial H_1}{\partial k_x}  | n \rangle \langle n |\frac{\partial H_1}{\partial k_y} |l \rangle   }{(\varepsilon_{l}-\varepsilon_{n})^2},
\end{eqnarray}
where $E_F$ is the Fermi energy, $\varepsilon_{l/n}$ and $|l/n \rangle$ are the corresponding eigenenergy and eigenstate of $H_1$, respectively.

For simplicity, in the following numerical calculations, we assume that the topological insulator film is spatially isotropic. The independent parameters are set to be: $A_{\alpha}=A=1.5$, $B_{\alpha}=B=1.0$, $D_{\alpha}=D=0.1$, $M=0.3$. $\alpha$ denotes $x,~y,~z$ and nearest hopping parameter $B$ is set as the energy unit. Our main results of this paper still hold for  anisotropic parameters or  more realistic parameters.~\cite{HJZhang}

\section{QAHE in the ideal neutral samples}\label{HalfFilling}
In this section, we focus on the topological phenomena in the ideal neutral samples. Physically, in our model, ideal neutral means that in the clean topological insulator film all  valence bands are occupied while all  conduction bands are unoccupied. Moreover,  magnetic doping does not bring in some extra carriers, and the system is not affected by the external environment.
\begin{figure}
\includegraphics [width=\columnwidth, viewport=31 33 757 592, clip]{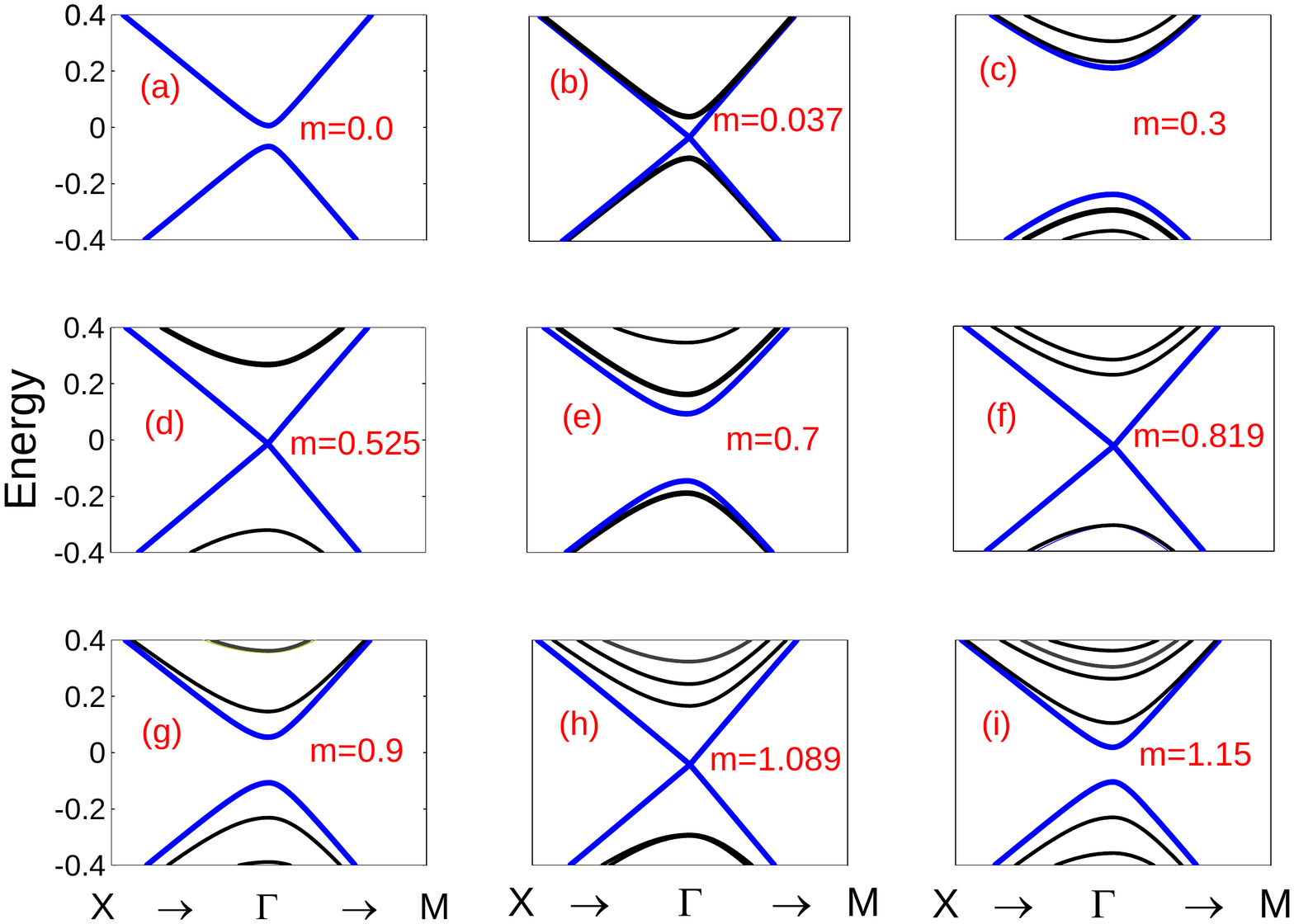}
\caption{(Color online)
The bulk band spectra along the high-symmetry lines of the magnetic element doped topological insulator film in the presence of different exchange field strength $m$=0(a), 0.037(b), 0.3(c), 0.525(d), 0.7(e), 0.819(f), 0.9(g), 1.089(h), 1.15(i). The layer-thickness is set to be $\rm N_z=12$.}\label{bulkbandsevolution}
\end{figure}

According to the definition of the topological order, any two insulating states are topologically equivalent only when they can adiabatically change into each other. Therefore, the classification of insulating state is highly related to the evolution of bulk energy spectrum. If bulk gap closes and reopens, the initial state and finial state cannot be adiabatically connected. Consequently, they belong to different topological phases and undergo a quantum phase transition. In our system, the time-reversal symmetry is broken due to the presence of  exchange field, which is different from  magnetic field in the conventional quantum Hall effect. In  realistic experiments, the exchange field strength could be tuned through controlling the doping concentration
of magnetic elements.

In Fig.~\ref{bulkbandsevolution}, we plot the bulk energy spectrum along the high symmetry lines for different exchange field strength $m$. Without the exchange field, the bulk band gap opens, which indicates an insulating state [see in panel (a)]. When the exchange field is introduced, the bulk gap gradually decreases along with the increasing of the exchange field $m$. As shown in panel (b), when the exchange field reaches the critical value $m_c=0.037$, the bulk band gap is completely closed. When the exchange field is further larger than $m_c=0.037$, the bulk band gap reopens [see in panel (c)]. This gap closing and reopening indicate a quantum phase transition, which has been pointed out in Ref.~[\onlinecite{RuiYu}].

The major finding of this article is that the bulk band gap is not monotonous but oscillates as a function of the exchange field strength $m$. For example, besides $m_c=0.037$, we found that there are other critical points (band gap closing and reopening), e.g. $m_c=0.525,~0.819,~1.089$. This indicates that there are more than two topologically different phases. From the critical points of view, one can see that in Fig.~\ref{bulkbandsevolution} there are five different phases.
\begin{figure}
\includegraphics [width=\columnwidth, viewport=51 33 777 562, clip]{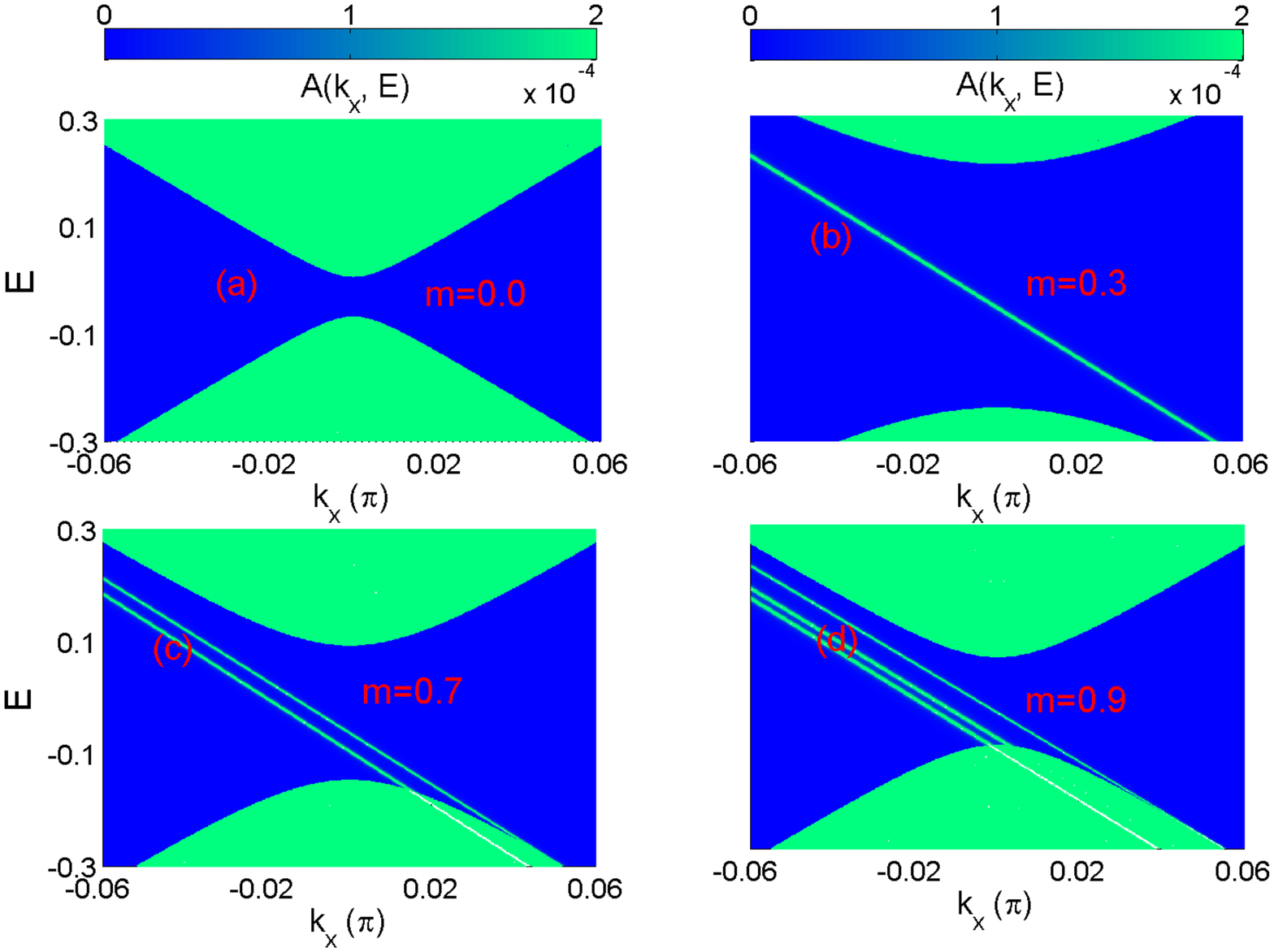}
\caption{(Color online)
Edge spectral functions $A(k_x,\omega)$ of the semi-infinite sample along y-direction for different exchange field strength $m$=0.0(a), 0.3(b), 0.7(c), 0.9(d). The sample thickness is set to be $\rm N_z=12$.}\label{edgeBands}
\end{figure}

One of the most striking characteristics of the nontrivial insulating states is the existence of  topologically protected gapless edge states, i.e. in  quantum Hall effect the number of  Chiral gapless edge states equals to the non-zero Chern number. Therefore, edge state analysis can be regarded as an efficient method to reveal the topology of the bulk states.~\cite{CLKane1} Figure~\ref{edgeBands} exhibits the edge spectral function ${\bf A}(k_x, E)$ for the first four topological phases with different exchange field strength being $m$=0.0(a),~0.3(b),~0.7(c),~0.9(d). Note that in all the figures of the edge spectral function ${\bf A}(k_x, E)$ the blue regime denotes  bulk gap, green regime denotes  bulk band states, and the lines represent  edge states at the boundary.

\begin{figure}
\includegraphics [width=\columnwidth, viewport=51 28 777 552, clip]{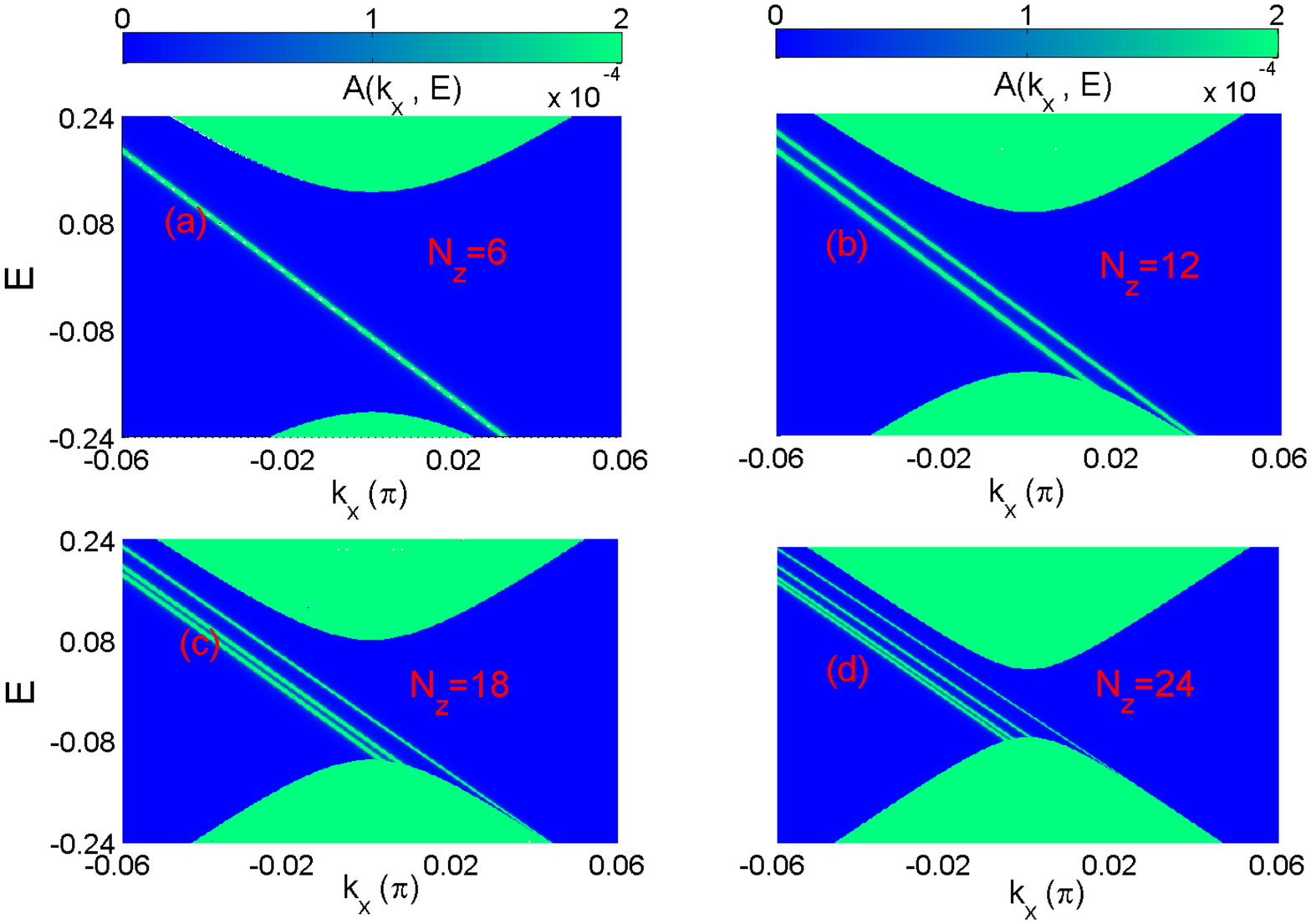}
\caption{(Color online)
Edge spectral functions $A(k_x,\omega)$ of the semi-infinite sample along y-direction for different sample thickness $\rm N_z$=6(a), 12(b), 18(c), 24(d). The exchange field strength is set to be $m=0.7$.  }\label{edgeThickness}
\end{figure}

In the absence of exchange field, one can notice that there is no edge state inside the bulk band gap in panel (a). This means that the system has the same topology as that of the vacuum and is considered as a trivial insulator. When the system enters the second topological phase, i.e. $0.037<m<0.525$, one chiral gapless edge state appears inside the bulk band gap~[see in panel (b) for $m=0.3$], indicating that the system becomes a nontrivial insulator with winding number $N=1$. According the relationship between  winding number of the edge state and the bulk Chern number $\mathcal{C}$,~\cite{YHatsugai} we claim that the system belongs to the $\mathcal{C}=1$ QAHE phase. After the system undergoes the second quantum phase transition, we observe that there are two chiral gapless edge states locating inside the bulk band gap [see in panel (c) for $m=0.7$], which labels that our system enters another QAHE phase with Chern number being $\mathcal{C}=2$. By continuously increasing the exchange field strength $m$, one can obtain $N=2,3,4,5...$ chiral edge states. In other words, the QAHE phase with various non-zero Chern numbers can be achieved in our system. Besides controlling the exchange field amplitude, we further find that the consecutive topological phase transitions can also happen through varying the sample thickness $\rm N_Z$, when the exchange field strength $m$ is larger than the gap controlling parameter $M$. This can be concluded in Fig.~\ref{edgeThickness} showing the evolution of the edge states with different  sample thickness $\rm N_Z$.

\begin{figure}
\includegraphics [width=\columnwidth, viewport=21 08 757 622, clip]{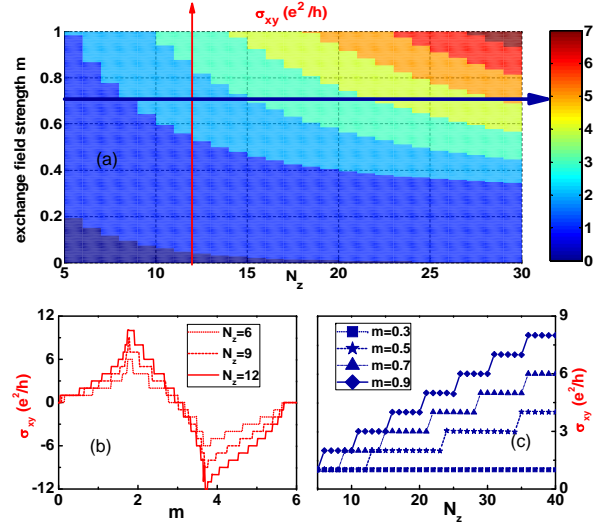}
\caption{(Color online)
(a) Phase diagram of the Hall conductance $\sigma_{xy}$ in the plane ($m$, $\rm{N_z}$). (b) and (c) plot the Hall conductance for fixed $\rm N_Z=$6/9/12 and $m$=0.3/0.5/0.7/0.9. The phenomena beyond the parameter space in panel (a) are also shown.}\label{HallConductance}
\end{figure}

To give a better understanding of this QAHE phases with various Chern numbers, we studied their topological properties by directly calculating the zero-temperature Hall conductance. In Fig.~\ref{HallConductance}, we calculated the Hall conductance $\sigma_{xy}$ as functions of the exchange field $m$ and the sample thickness $\rm N_Z$. In panel (a), the phase diagram of the Hall conductance $\sigma_{xy}$ in  ($\rm N_Z$, $m$) plane is plotted, and different colors are used to specify kinds of Hall plateaus. One can find that the Hall conductance varies as functions of both the exchange field and  sample thickness. To be more specific, in panel (b) we show the Hall conductance as function of the exchange field at fixed sample thickness $\rm {N_Z}=6~,9,~12$; in panel (c) we plot the Hall conductance as a function of the sample thickness at fixed exchange field $m$=0.3,~0.5,~0.7, and~0.9. Since the topological invariant Chern number in the QAHE is identical to the Hall conductance (in units of $e^2/h$), therefore each separated QAHE phase and the resulting phase boundaries can be easily determined. Figure~\ref{HallConductance} is the central result of this article. The obtained QAHE phases manifest the following features: (i) Multiple discrete QAHE phases exist in our system with phase transitions directly from one to another by varying the exchange field strength or  sample thickness; (ii) In contrast to the reported QAHE proposals, the Chern number in our proposal can be very high and the highest Chern number is comparable to the total layers $\rm N_Z$; (iii) the Hall conductance $\sigma_{xy}$ is not monotonous as a function of the exchange field strength $m$. For instance, panel (b) can be divided into three regions where  Hall conductance $\sigma_{xy}$ is stepped by 1, -2, and 1 along with the increasing of  exchange field strength $m$; (iv) We emphasize that the relationship between $\sigma_{xy}$ and $m$ in our model resembles that between $\sigma_{xy}$ and the external magnetic field in a conventional quantum Hall effect.

So far, we have showed the existence of the tunable Chern number QAHE phases and demonstrated their topological properties in the neutral samples. In the following, we will move to the physical mechanism leading to such phenomena. Compared to the previous two-dimensional QAHE models,~\cite{Haldane,ZHQiao,RuiYu,XLQi}, our studied model is quasi-three dimensional. However, to satisfy the condition of requiring the vanishing wave-function at the regions $z<0$ and $z> {\rm N_z} $, the wave vector $k_z$ in Z-direction should take real discrete values with their magnitude approach $k_z =\frac{n\pi}{{\rm N_z}} (n=1,2,3,...,{\rm N_z})$ or one imaginary value\cite{SQShen1, addnote1}. The latter case leading to the $\mathcal{C}=1$ QAHE phase reported in Ref.~[\onlinecite{RuiYu}]. We will focus on the former one. In the momentum space, the Hamiltonian of our system can be written as $H(k)=\sum_{k_z} H_{k_z} (k_x,k_y)$ with $k_x,~k_y$ being the  momenta along $x$ and $y$ directions and $k_z$ taking some concrete real constants. In this way, our model can be regarded as the combination of a series of two-dimensional square lattice models described by $H_{k_z}(k_x,k_y)$. Moreover, $H_{k_z}(k_x, k_y)$ can be block-diagonalized and the new diagonal Hamiltonian $\overline{H_{k_z}}(k_x,k_y)$ is expressed as:
\begin{eqnarray}
\overline{H_{k_z}}(k_x,k_y)&=&  \varepsilon_{k_z}(k_x,k_y) + \left(
                                              \begin{array}{cc}
                                              h_{k_z}^1(k_x,k_y)  & O \\
                                                O &  h_{k_z}^2(k_x,k_y) \\
                                              \end{array}
                                            \right)  \nonumber\\
h_{k_z}^1(k_x,k_y) &=&  d_z^{1} \sigma_z +d_x \sigma_x +d_y \sigma_y  \nonumber\\
h_{k_z}^2(k_x,k_y) &=& d_z^{2} \sigma_z +d_x \sigma_x +d_y \sigma_y \label{equation7}
\end{eqnarray}
where the parameters $\varepsilon_{k_z}$, $d_x$, $d_y$, $d^1_z$, $d^2_z$ are written as:
\begin{eqnarray}
\varepsilon_{k_z}(k_x,k_y) &=& -3 D +D (\cos k_x +\cos k_y +\cos k_z) \nonumber\\
d_x &=& A \sin k_x  \nonumber\\
d_y &=& A \sin k_y  \nonumber\\
d_z^{1}(k_x,k_y) &=& - \sqrt{ [M_{k_z}(k_x,k_y)]^2 + A^2 \sin ^2k_z} +m \nonumber\\
d_z^{2}(k_x,k_y) &=& \sqrt{ [M_{k_z}(k_x,k_y)]^2 + A^2 \sin ^2k_z} +m \nonumber\\
M_{k_z}(k_x,k_y)&=& M-3B+B(\cos k_x +\cos k_y +\cos k_z) \nonumber\\
\end{eqnarray}

One can find that both  two block Hamiltonian $h_{k_z}^1$ and $h_{k_z}^2$ are the Haldane's model that is similar to the model described in Ref.~[{\onlinecite{XLQi}}]. Only when $d_z^{1} (d_z^{2})$ can change its sign in the first Brillouin zone, the two bands of $h_{k_z}^1 (h_{k_z}^2)$ become inverted to result in  QAHE with $\mathcal{C}=+1 (-1)$ Chern number.~\cite{XLQi} For positive exchange field strength $m$, $h_{k_z}^2$ is a trivial insulator since $d_z^2$ cannot change its sign. In contrast, $d_{z}^{1}$ can change its sign and consequently $h_{k_z}^1$ describes the QAHE for certain positive exchange field strength $m$. Combining the results for $h_{k_z}^1$ and $h_{k_z}^2$, the total Chern number of $\overline{H_{k_z}}$  in the different zone of the exchange field strength $m$ can be expressed as:~\cite{XLQi}
\begin{eqnarray}
C= \begin{cases} 0 & 0< m<m_{c1}(k_z) \ \ or \ \ m > m_{c3} (k_z)  \\ 1 &  m_{c1}(k_z) < m < m_{c2} (k_z)  \\ -1 & m_{c2}(k_z) < m < m_{c3}(k_z)
\end{cases}\label{equation9}
\end{eqnarray}
with the constants $m_{c1},~m_{c2},~m_{c3}$ being:
\begin{eqnarray}
m_{c1}(k_z) &=& \sqrt{(M-B +B \cos k_z )^2 +A^2 \sin^2 k_z} \nonumber\\
m_{c2}(k_z) &=& \sqrt{(M-3B +B \cos k_z)^2 +A^2 \sin^2 k_z} \nonumber\\
m_{c3}(k_z) &=& \sqrt{(M-5B +B \cos k_z)^2 +A^2 \sin^2 k_z}. \label{equation10}
\end{eqnarray}

The Eqs.~\ref{equation7}-\ref{equation9} are the main mechanism of the tunable Chern number QAHE phases. The obtained results in the numerical simulations can be explained by such mechanism. In the following, we will demonstrate how it is applied to the continuously increasing exchange field strength $m$. Since $k_z$ takes a series of discrete values, the constants $m_{c1},~m_{c2},~m_{c3}$ also take a series of discrete values. When $m$ exceeds $m_{c1}$, the corresponding $\overline{H_{k_z}}$ transitions from the trivial insulator to the $\mathcal{C}=1$ QAHE phase. And consequently the quantized Hall conductance $\sigma_{xy}$ is increased to ${e^2}/{h}$. $\sigma_{xy}$ will keep as a constant until $m$ exceeds another $m_{c1}$. Therefore, the quantum Hall plateaus appear in Fig.~\ref{HallConductance}. As shown in panel (b), when  Hall conductance $\sigma_{xy}$ reaches its maxima, it will begin to drop ${2e^2}/{h}$ due to QAHE transition from $\mathcal{C}=1$ to $\mathcal{C}=-1$ for one $\overline{H_{k_z}}$. For even larger $m$ exceeding $m_{c3}$, $\overline{H_{k_z}}$ transitions from $\mathcal{C}=-1$ QAHE phase to a trivial insulator. Consequently, $\sigma_{xy}$ shows quantized plateaus with interval of ${e^2}/{h}$ at the right-hand-side of panel (b).

Next, let us explain the phenomena in the panel (c). According to Eqs.~(\ref{equation9}) and (\ref{equation10}), $m_{c1}$ blow the fixed $m$ will contribute one Chern number. Though $k_z$ and $m_{c1}$ vary for different sample thickness $\rm N_z$, the numbers of discrete $m_{c1}$  below $m$ will not change until $\rm N_z$ exceeds the critical thickness. This explains why the Hall conductance also shows quantized plateaus in panel (c). From Eq.~(\ref{equation10}) one can obtain that the minimum of $m_{c1}$ approaches the gap controlling parameter $M$. Therefore, in order to observe the higher Chern number QAH state, the exchange field strength should be larger than $M$. This can be verified in our numerical simulations, e.g. in  panel (c) only the $\mathcal{C}=1$ QAHE is obtained when $m$ is identical to gap parameter $M$.

\begin{figure}
\includegraphics [width=\columnwidth, viewport=11 28 757 602, clip]{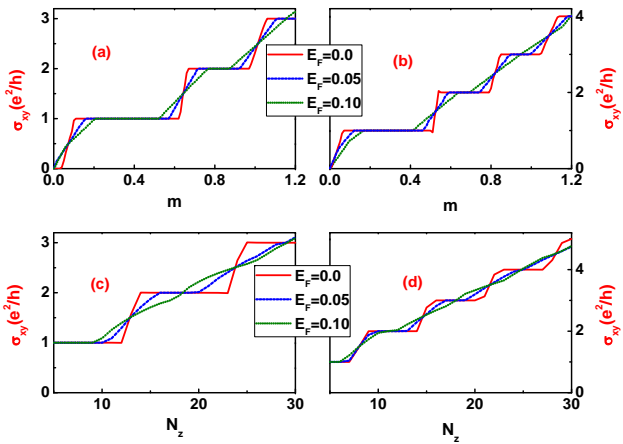}
\caption{(Color online)
(a)-(b): The Hall conductance $\sigma_{xy}$ as a function the exchange field strength $m$ under different Fermi energies $E_F=0.00,~0.05,~0.10$ at fixed sample thickness ${\rm N_z}=9$(a) and ${\rm N_z}=12$(b), respectively. (c)-(d): The Hall conductance $\sigma_{xy}$ as a function the layer thickness $m$ under different Fermi energies $E_F=0.00,~0.05,~0.10$ at fixed exchange field ${m}=0.5$(a) and $m=0.7$(b), respectively.  }\label{ElectronFermi}
\end{figure}

\section{Robustness of the tunable QAHE in the realistic samples}\label{RealisticSamples}
In the above section, we study the topological phenomena in the ideal neutral samples. However, in the realistic materials such ideal conditions can hardly be satisfied. In general, there are two kinds of inevitable impurity effects. First, the impurities may bring in some extra carriers that make the Fermi energy shift from the bulk band gap. Second, the imperfect especially nonuniform doping may induce the disorder effects leading to the magnetic or nonmagnetic scattering. In the following, the influence of these two effects on the high Chern number QAHE will be addressed.

We first concentrate on the case that  impurities provide some extra carriers. Fortunately, one can overcome this difficulty by utilizing an external gate to fix the  Fermi energy into  band gap [see as figure 1]. In experiment, the Fermi energy can be tuned by the gate voltage.
In Fig.~\ref{ElectronFermi}, the Hall conductance $\sigma_{xy}$ as the function of  exchange field strength $m$ or sample thickness ${\rm N_z}$ for different fixed Fermi energies $E_F$ is plotted. The presence of the quantized plateaus prove that the high Chern number QAHE phase can still hold to some degree. Nevertheless, a regime with continuously varying Hall conductance exists between two consecutive quantized Hall plateaus, which indicates that the metallic phases emerge between two separated QAHE phases. For the fixed Fermi energy shift, the metallic phases will easily show up in the high Chern number QAHE regimes with strong exchange field strength $m$ or large sample thickness ${\rm N_z}$.

As discussed in the above section, our model can be considered as a series of two-dimensional Haldane model [see in Eq.~(\ref{equation7})] that is characterized by the wave vector $k_z$. Obviously, in order to obtain the quantized Hall conductance, the Fermi energy should locate inside the bulk band gaps of all the Haldane models, i.e. from Eq. (8) and (9) all  possible discrete $k_z$ should satisfy the following relationship:
\begin{eqnarray}
\varepsilon_{k_z}(0,0)-|d_z^{1}(0,0)|< E_F < \varepsilon_{k_z}(0,0)+|d_z^{1}(0,0)| \label{criteria}
\end{eqnarray}
If  Eq.~(\ref{criteria}) is not satisfied, the metallic phase will appear. A more detailed analysis of Eq.~(\ref{criteria}) shows that in order to find the high Chern number QAHE in the realistic materials, the smaller $m$ or thinner ${\rm N_z}$ as well as the lower  Fermi energy shift are required. This analytical conclusion agrees well with our numerical simulations.

\begin{figure}
\includegraphics [width=\columnwidth, viewport=36 48 731 577, clip]{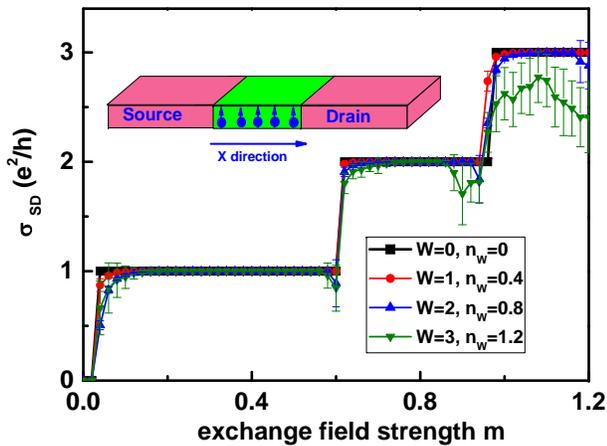}
\caption{(Color online)
The longitudinal conductance $\sigma_{SD}$  as a function of the exchange field strength $m$ under different white noise disorder strength $W$ and
the nonuniform doping factor $n_W$ combinations. The error bar is used to denote the conductance fluctuation $\delta \sigma_{SD}$. The Fermi energy is set to be $E_F=0.05$. Inset: the schematic plot of the two-terminal device. Disorders are only considered in the central scattering regime, which is modeled by a ${\rm N_x}\times N_y \times N_z=30\times40\times9$ cubic lattice models.}\label{two-terminalSetup}
\end{figure}

Secondly, let us discuss the disorder effect. Since  disorder breaks the translational symmetry, it is difficult to study its effect on the topological properties using  energy band theory. However, through analyzing the resulting transport properties,  disorder effect on the bulk topological properties could be estimated. In the following, we apply our tight-binding model to a two-terminal device and study its transport properties using the nonequilibrium Green's function method. Here we will not introduce the calculating method in detail which could be found in Ref.~[\onlinecite{HJIANG}]. Our considered model is illustrated in the inset of Fig.~\ref{two-terminalSetup}. The disorders are only considered in the central scattering region. The source and drain are perfect semi-infinite leads which will not bring any redundant scattering at the interfaces between the leads and central region. In general, there are two kinds of disorders existing in a real system: the white noise and the nonuniform doping. In our model, the white noise is modeled by the on-site disorder energy $W_i \sigma_0 \otimes \tau_0 $ with $W_i$  uniformly distributed in the interval of $[-\frac{W}{2},\frac{W}{2}]$, where $W$ measures the strength of  disorder. The nonuniform magnetic doping is modeled by the uniform background which produces  on-site zeeman energies $m \sigma_z \otimes \tau_0 $ and the exchange field at different sites fluctuate leading to $m n_i \sigma_z \otimes \tau_0 $ on each site. Here $n_i$ is uniformly distributed in the range $[-\frac{n_W}{2},\frac{n_W}{2}]$ with $n_W$ being the nonuniform doping factor.

Figure~\ref{two-terminalSetup} plots the longitudinal conductance $\sigma_{SD}$ and the corresponding conductance fluctuation $\delta \sigma_{SD}$ versus the exchange field strength $m$ for different combinations ($W,~n_W$)=(0, 0), (1, 0.4), (2, 0.8), and (3, 1.2). In absence of disorder, $\sigma_{SD}$ shows perfect quantized conductance plateaus (see the squared-symbol curve). When  disorder is present, we found that around the regimes close to the quantum phase transition (e.g. $m \in [0.94,1.06]$) the conductance is easy to be destroyed indicating that the disorder results in the metallic phase permitting the state being backscattered. The obtained metallic region is in excellent consistence with the results from the Hall conductance calculation with the Fermi-energy shift. While in the central regime of each plateau (e.g. $m \in [0.74, 0.86]$) the longitudinal conductance $\sigma_{SD}$ remains quantized with vanishing conductance fluctuations $\delta \sigma_{SD}$ even when the white noise disorder strength approaches $W=3$ and nonuniform factor reaches $n_W=1.2$. Since the existence of chiral edge states is the only possible mechanism to interpret the absence of backscattering, one can conclude that the high Chern number QAHE phases are robust against strong white noise and nonuniform magnetic doping.~\cite{addnote2}

\section{Discussion and Conclusion}\label{Summary}
One challenge of obtaining the high Chern number QAHE phase is that the exchange field strength $m$ should be larger than the bulk band gap amplitude of  topological insulator. This condition is hardly to be realized in $Bi_2Se_3$ and many other obtained materials. However, with the rapid development of this field\cite{SuYangXu}, it is highly possible to find some narrow-gap topological insulators in the near future. In addition, we want to emphasize that the sign of  gap controlling parameter $M$ is not important. In other words, one can obtain the tunable high Chern number QAHE in  semi-metal or narrow-gap normal semiconductor though the QAHE phase arisen from the surface state is difficult to be realized.

Moreover, we want to discuss the relationship between the QAHE and the extended Haldane model. In the main text, we have claimed that the extended Haldane model can be used to explain the emergence of the quantized Hall plateaus. However, in the realistic materials there are several features that might destroy certain characteristics of the system, thus deviates the scope of the extended Haldane model. For example, (i) when the topological insulator is doped with some magnetic impurities,  Lande $g$ factors of all the occupied orbitals might be slightly different, thus the mass-term in the Hamiltonian has diverse values and the Hamiltonian cannot be deduced to the Haldane model; (ii)  impurities may destroy the inversion symmetry, which leads to a much more complicated situation that cannot be captured by the Haldane model. Fortunately, in our numerical simulations we have considered this two important features, and our results indicate that the tunable high Chern number QAHE plateaus are very robust against these two kinds of violations. This robustness can be attributed to the fact that the discrete finite $k_z$ protects the system from small disturbance and the stability of QAHE under disorders should be very significant in the experimental realization.

In summary, using the tight-binding method, we demonstrate that tunable Chern number QAHE can be in principle observed through doping magnetic elements in the topological insulator film. The perfect quantum plateaus in the ($m$, ${\rm N_z}$) plane exist in the ideal neutral samples. The topological properties
of the tunable high Chern number QAHE are discussed via the evolution of  bulk and edge states as well as the direct Hall conductance calculation. Moreover, the physical origins of the tunable Chern number QAHE phase are explained using the extended Haldane model. Further, we also study the robustness of the tunable Chern number QAHE at fixed Fermi-energy and nonmagnetic/magnetic disorders. Our theoretical prediction should shed some light on the searching of high Chern number QAHE materials.

\section{Acknowledgement}
We are grateful to Junren Shi, Zhong Fang, Ying Ran, Di Xiao, Xi Chen, Yugui Yao and X.C. Xie for many helpful discussions. H.J. was supported by China Postdoctroal Science Foundation (Grant No. 20100480147) and 985 program of Peking University. Z.Q. was supported by NSF (Grant No. DMR 0906025)
and Welch Foundation~(Grant No. F-1255). Q.N. was supported by
DOE~(Grant No. DE-FG03-02ER45958, Division of Materials Science and
Engineering) and Texas Advanced Research Program.


\begin{thebibliography}{99}
\bibitem{Klitzing}
K.v. Klitzing, G. Dorda and M. Pepper, Phys. Rev. Lett. {\bf 45}, 497 (1980).

\bibitem{DCTUSI}
D.C. Tusi, H.L. Stormer, and A.C. Gossard, Phys. Rev. Lett. {\bf 48}, 1559 (1982).

\bibitem{RBLauglin}
R.B. Laughlin, Phys. Rev. Lett. {\bf 50}, 1395 (1983).

\bibitem{Thouless}
D.J. Thouless, M. Kohmoto, M.P. Nightingale and M.D. Nijs, Phys. Rev. Lett. {\bf 49}, (1982); M. Kohmoto, Ann. Phys. (N.Y.) {\bf 160}, 343 (1985).

\bibitem{Haldane}
F.D.M. Haldane, Phys. Rev. Lett. {\bf 61}, 2015 (1988).

\bibitem{CXLiu}
C.-X. Liu, X.-L. Qi, X. Dai, Z. Fang, and S.-C. Zhang, Phys. Rev. Lett. {\bf 101}, 146802 (2008).

\bibitem{Nagaosa}
M. Onoda {\em et al.}, Phys. Rev. Lett. {\bf 90}, 206601 (2003).

\bibitem{ZHQiao}
Z.H. Qiao, S.A. Yang, W.X. Feng, W.-K. Tse, J. Ding, Y.G. Yao, J. Wang, and Q. Niu, Phys. Rev. B {\bf 82}, 161414(R) (2010); W.-K. Tse, Z.H. Qiao, Y.G. Yao, A.H. MacDonald, and Q. Niu, Phys. Rev. B, {\bf 83}, 155447 (2011);  T.-W. Chen, Z.-R. Xiao, D.-W. Chiou, and G.-Y. Guo, arXiv: 1103.4083; J. Ding, Z.H. Qiao, W.X. Feng, Y.G. yao, and Q. Niu, arXiv:1108.6235v1 (2011).

\bibitem{Nagaosa1}
K. Ohgushi, S. Murakami, and N. Nagaosa, Phys. Rev. B {\bf 62}, 6065 (2000); K. Sun, H. Yao, E. Fradkin, and S.A. Kivelson, Phys. Rev. Lett. {\bf 103}, 046811 (2009); J. Wen, A. Ruegg, C.C.J. Wang, and G.A. Fiete Phys. Rev. B {\bf 82}, 075125 (2010).

\bibitem{ZYZhang}
Z.-Y. Zhang, J. Phys.: Condens. Matter, {\bf 23}, 365801 (2011).

\bibitem{CJWu}
C.J. Wu, Phys. Rev. Lett. {\bf 101}, 186807 (2008); X.-J. Liu, X. Liu, C.J. Wu, and J. Sinova, Phys. Rev. A {\bf 81}£¬033622 (2010); M. Zhang, H.-H. Hung, C.W. Zhang, and C.J. Wu, Phys. Rev. A, {\bf 83}, 023615 (2011); Y.P. Zhang, and C.W. Zhang, Phys. Rev. B, {\bf 84}, 085123(2011).

\bibitem{CLKane1}
M.Z. Hasan and C.L. Kane, Rev. Mod. Phys {\bf 82}, 3045 (2010); X.-L. Qi and S.-C. Zhang  Rev. Mod. Phys {\bf 83}, 1057 (2011);
J.E. Moore, Nature(London), {\bf 464}, 194 (2010).

\bibitem{CLKane2}
C.L. Kane and E.J. Mele, Phys. Rev. Lett. {\bf 95} 226801 (2005).

\bibitem{SCZhang1}
B.A. Bernevig, T.L. Hughes, and S.-C. Zhang, Science, {\bf 314}, 1757 (2006).

\bibitem{SCZhang2}
M. K$\ddot{o}$nig, S. Wiedmann, C. Br$\ddot{u}$ne, A. Roth, H. Buhmann, L. W. Molenkamp, X.-L. Qi and S.-C. Zhang, Science {\bf 318}, 766 (2007).

\bibitem{Fertig}
E. Prada, P. San-Jose, L. Brey, H.A. Fertig, Solid State Commun. {\bf 151}, 1075 (2011)

\bibitem{ZHQiao1}
Z.H. Qiao, W.-K. Tse, H. Jiang, Y.G. Yao, and Q. Niu, arXiv:1109.1131v1 (2011).

\bibitem{RQWu}
C. Weeks, J. Hu, J. Alicea, M. Franz, and R.Q. Wu, Phys. Rev. X {\bf 1}, 021001 (2011).

\bibitem{RRDu}
C.-X. Liu, T. L. Hughes, X.-L. Qi, K. Wang, and S.-C. Zhang, Phys. Rev. Lett {\bf 100}, 236601, (2008);  I. Knez, R.-R. Du, and G. Sullivan, Phys. Rev. Lett {\bf 107}, 136603 (2011).

\bibitem{YGYao}
C.-C. Liu, W.X. Feng, and Y.G. Yao, Phys. Rev. Lett {\bf 107}, 076802 (2011); C.-C. Liu, H. Jiang and Y.G. Yao
arXiv:1108.2933v1(2011).

\bibitem{LiangFu}
L. Fu and C.L. Kane, Phys. Rev. B  {\bf 76}, 045302 (2007); L. Fu, C.L. Kane, and E.J. Mele,
Phys. Rev. Lett. {\bf 98}, 106803 (2007).

\bibitem{HJZhang}
H.J. Zhang, C.-X. Liu, X.-L. Qi, X. Dai, Z. Fang and S.-C. Zhang,
Nat. Phys. {\bf 5}, 438 (2009).

\bibitem{Dixiao}
H. Lin, L.A. Wray, Y.Q. Xia, S.Y. Xu, S. Jia, R.J. Cava, A. Bansil and M.Z. Hasan,
Nat. Mat. {\bf 9}, 546 (2010); S. Chadov, X.L. Qi,  J. K$\ddot{u}$bler, G.H. Fecher, C. Felser, and S.-C. Zhang,  Nat. Mat. {\bf 9}, 541 (2010); D. Xiao, Y. Yao, W.F. Feng, J. Wen, W.G. Zhu, X.-Q. Chen, G.M. Stocks and Z.Y. Zhang,  Phys. Rev. Lett. {\bf 105}, 096404 (2010).

\bibitem{ZhongFang}
W. Zhang, R. Yu, W.X. Feng, Y.G. Yao, H.M. Weng, X. Dai, and Z. Fang, Phys. Rev. Lett. {\bf 106}, 156808 (2011).

\bibitem{Yansun}
Y. Sun, X.-Q. Chen, S. Yunoki, D.Z. Li, and Y.Y. Li, Phys. Rev. Lett. {\bf 105}, 216406 (2010).

\bibitem{Virot}
F. Virot, R. Hayn, M. Richter, and J. van den Brink, Phys. Rev. Lett. {\bf 106}, 235806 (2011).

\bibitem{DHsieh}
D. Hsieh, D. Qian, L. Wray, Y. Xia, Y. S. Hor, R. J. Cava and M. Z. Hasan, Nature {\bf 452}, 970 (2008).

\bibitem{Xia}
Y. Xia D. Qian, D. Hsieh, L. Wray, A. Pal, H. Lin, A. Bansil, D. Grauer, Y. S. Hor, R. J. Cava and M. Z. Hasan
Nat. Phys. {\bf 5}, 398 (2009); Y. L. Chen, J. G. Analytis, J.-H. Chu, Z. K. Liu, S.-K. Mo, X. L. Qi, H. J. Zhang, D. H. Lu, X. Dai, Z. Fang, S. C. Zhang, I. R. Fisher, Z. Hussain and Z.-X. Shen, Science {\bf 325}, 178 (2009).

\bibitem{Sato}
T. Sato, K. Segawa, H. Guo, K. Sugawara, S. Souma,
T. Takahashi, and Y. Ando, Phys. Rev. Lett. {\bf 105}, 136802 (2010);
K. Kuroda, M. Ye, A. Kimura1, S. V. Eremeev, E. E. Krasovskii, E. V. Chulkov, Y. Ueda, K. Miyamoto, T. Okuda, K. Shimada, H. Namatame, and M. Taniguchi, Phys. Rev. Lett. {\bf 105} 216406 (2010); Y.L. Chen, Z.K. Liu, J. G. Analytis, J.-H. Chu, H.J. Zhang, B.H. Yan, S.-K. Mo, R.G. Moore, D.H. Lu, I.R. Fisher, S.-C. Zhang, Z. Hussain, and Z.-X. Shen, Phys. Rev. Lett.  {\bf 105}, 266401 (2010).

\bibitem{RuiYu}
R. Yu, W. Zhang, H.-J. Zhang, S.-C. Zhang, X. Dai, Z. Fang, Science {\bf 329}, 61 (2010).

\bibitem{QKXue}
C.-Z. Chang, J.-S. Zhang, M.-H. Liu, Z.-C. Zhang, X. Feng, K. Li, L.-L. Wang, X. Chen, X. Dai, Z. Fang, X.-L. Qi, S.-C. Zhang, Y.Y. Wang, K. He, X.-C. Ma, Q.-K. Xue, arXiv:1108.4754 (2011).

\bibitem{ChangKai}
Q. Liu, C. X. Liu, C. K. Xu, X. L. Qi, S. C. Zhang, Phys. Rev. Lett. 102, 156603 (2009);
J.-J. Zhu, D.-X. Yao, S.-C. Zhang, and K. Chang, Phys. Rev. Lett. {\bf 106}, 096201 (2011).

\bibitem{YLChen}
Y.L. Chen, J.-H. Chu, J.G. Analytis, Z.K. Liu, K. Igarashi, H.-H. Kuo, X.-L. Qi, S.K. Mo, R.G. Moore, D.H. Lu, M. Hashimoto, T. Sasagawa, S.-C. Zhang, I.R. Fisher, Z. Hussain, and Z.X. Shen, Science, {\bf 6}, 659 (2010);L. A. Wray, Su-Yang Xu, Yuqi Xia, D. Hsieh, A. V. Fedorov, Hsin Lin, A. Bansil, Yew San Hor, R. J. Cava, and M. Z. Hasan, Nature Physics, {\bf 7}, 32 (2011); Y. S. Hor, P. Roushan, H. Beidenkopf, J. Seo, D. Qu, J. G. Checkelsky, L. A. Wray, D. Hsieh, Y. Xia, S.-Y. Xu, D. Qian, M. Z. Hasan, N. P. Ong, A. Yazdani, and R. J. Cava, Phys. Rev. B {\bf 81}, 195203 (2009).


\bibitem{SQShen}
R.-L. Chu, J.R. Shi, and S.-Q. Shen, Phys. Rev. B {\bf 84}, 085312 (2011).

\bibitem{HMGuo}
H.-M. Guo, G. Rosenberg, G. Refael, and M. Franz, Phys. Rev. Lett. {\bf 105}, 216601 (2010).

\bibitem{Lopez}
M.P.Lopez Sancho, J.M.Lopez Sancho, and J. Rubio, J. Phys. F: Met. Phys. {\bf 14} , 1205(1984); {\bf 15}, 851 (1985).

\bibitem{YHatsugai}Y. Hatsugai, Phys. Rev. B {\bf 48}, 11851 (1993).

\bibitem{XLQi}
X.-L. Qi, Y.S. Wu, and S.-C. Zhang, Phys. Rev. B {\bf 74}, 085308 (2006).

\bibitem{SQShen1}
H.-Z. Lu, W.-Y. Shan, W. Yao, Q. Niu, and S.-Q. Shen, Phys. Rev. B {\bf 81}, 115407 (2011).


\bibitem{addnote1}
$k_z$ taking the imaginary value means that the mode is a surface state. The coupling between the top and bottom surface modes opens a bulk gap. The closing and reopening of such a gap by the exchange field convert the normal insulator to $\mathcal{C}=1$ QAHE phase transition. For more details, please see in Ref.~[\onlinecite{RuiYu}].

\bibitem{HJIANG}
H. Jiang, L. Wang, Q.-F. Sun, and X.C. Xie, Phys. Rev. B {\bf 80}, 165316 (2009).


\bibitem{addnote2}
Actually, whether the QAHE proposed in Ref.~[\onlinecite{RuiYu}] can be realized in experiment is still in debate, since the first principles calculation method can hardly capture the nonuniform doping effect which is inevitably present in the realistic materials. Our work gives a positive answer to this debate.

\bibitem{SuYangXu}
S.-Y. Xu, Y. Xia, L. A. Wray, S. Jia, F. Meier, J. H. Dil, J. Osterwalder,
B. Slomski, A. Bansil, H. Lin, R. J. Cava, M. Z. Hasan. Science {\bf 332}, 560 (2011);
T. Sato, K. Segawa, K. Kosaka, S. Souma, K. Nakayama, K. Eto, T. Minami, Y. Ando
and T. Takahashi1, Nat. Phys. doi:10.1038/nphys2058.

\bibitem{addnote3}
In the final preparation of this article, we become awared of a related work in arXiv:1110.1939v1 by G.Y. Cho using the continuum model. The main result of our work has been posted in the conference ``The 4th International Workshop on Emergent Phenomena in Quantum Hall Systems (June 23 to 26, 2011 at Peking University, Beijing, China)''.
\end{thebibliography}
\end{document}